\documentclass{PoS}

\title{Predicting radiative B decays to vector mesons in holographic QCD}

\ShortTitle{Predicting radiative B decays to vector mesons in holographic QCD}

\author{Mohammad AHMADY\\
       Mount Allison University \\
       E-mail: \email{mahmady@mta.ca}}

\author{\speaker{Ruben SANDAPEN}%
         \thanks{This research is supported by NSERC. RS also thanks Mount Allison University and Universit\'e de Moncton for funding. }\\
        Universit\'e of Moncton $\&$ Mount Allison University \\
        E-mail: \email{ruben.sandapen@umoncton.ca}}

\abstract{We predict observables in rare radiative $B$ decays to vector mesons using holographic AdS/QCD Distribution Amplitudes (DAs) for the vector meson. We find that end-point divergences can be avoided when computing power-suppressed contributions in the heavy quark limit. The results reported here can be found in \cite{Ahmady:2012dy} and \cite{Ahmady:2013cva}. }

\FullConference{XXI International Workshop on Deep-Inelastic Scattering and Related Subjects\\
                 22-26 April, 2013\\
                 Marseilles, France}
\usepackage{amsmath}
\usepackage{bm}

\def\d{\mathrm{d}}
\begin{document}

\section{Introduction}
The standard theoretical framework for computing rare radiative $B$ decays is QCD factorization (QCDF) \cite{Bosch:2001gv}.   QCDF is the statement that to leading power accuracy in the heavy quark limit, the matrix element of the effective weak Hamiltonian operators factorizes into perturbatively calculable kernels and non-perturbative but universal quantities namely the $B \to V$ transition form factor, the meson decay constants  and their leading twist DAs.  In a standard notation \cite{Bosch:2001gv,Ball:2006eu}, we have
\begin{equation}
\langle V(P,e_T) \gamma (q,\epsilon ) | Q_i | \bar{B} \rangle = [ F^{B\rightarrow V} T_i^I  + \int_0^1 \d \zeta\; \d z \; \Phi_B(\zeta) T_i^{II}(\zeta,z)  \phi^{\perp}_{V} (z)] \cdot \epsilon   + \mathcal{O}(\Lambda_{\mbox{\tiny{QCD}}}/m_b)  \;.
\label{factorization}
\end{equation}
 The second term is a convolution of the perturbatively computable kernels $T_i^{II}$ with the non-perturbative leading twist DA of the $B$ meson, $\Phi_B(\zeta)$, and that of the vector meson, $\phi_{V}^{\perp}(z)$. 

The predictive power of QCDF is limited by two sources of uncertainty : firstly by the uncertainties associated with all non-perturbative quantities which we refer to as hadronic uncertainties and secondly by power corrections to the leading power contribution given by equation \eqref{factorization}.  Traditionally, the vector meson DAs are obtained using QCD Sum Rules which predict their moments at a starting low scale of $1$ GeV. The DAs are then reconstructed as a truncated Gegenbauer expansions and can then be evolved perturbatively to any desired higher scale. The computation of power corrections with Sum Rules DAs can be problematic because they involve convolution integrals that do not always converge. We refer to this as the end-point divergence problem \cite{Pecjak:2008gv}. 

In this contribution, we report that to leading power accuracy, predictions generated with alternative AdS/QCD DAs agree with those predicted with standard Sum Rules DAs \cite{Ball:2007zt} but that beyond leading power accuracy, AdS/QCD DAs offer the advantage of avoiding the end-point divergence problem. Furthermore, the AdS/QCD predictions agree with experiment.

\section{Holographic AdS/QCD wavefunctions for vector mesons}
\begin{figure}
\begin{center}
 \includegraphics[height=.15\textheight]{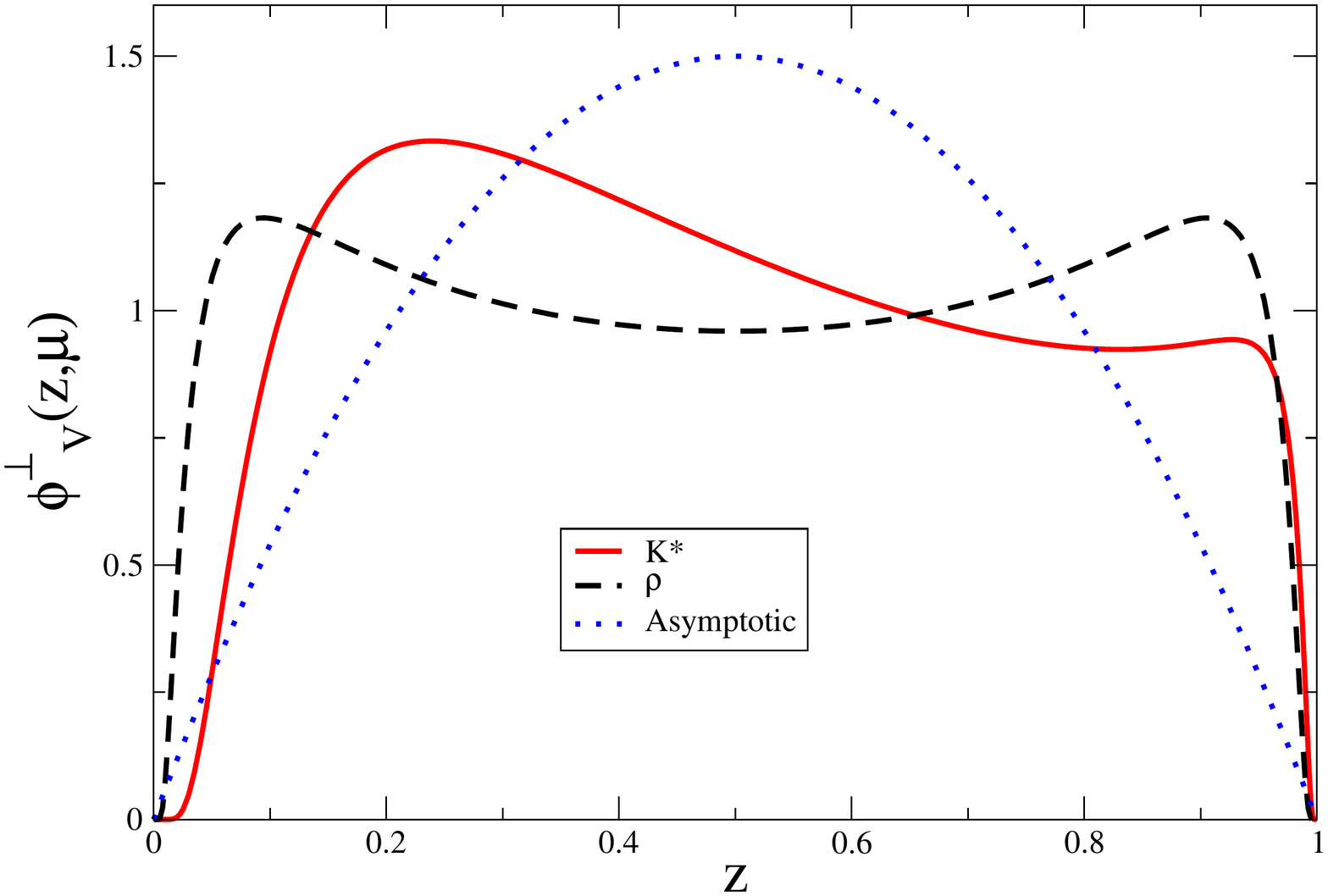}\includegraphics[height=.15\textheight]{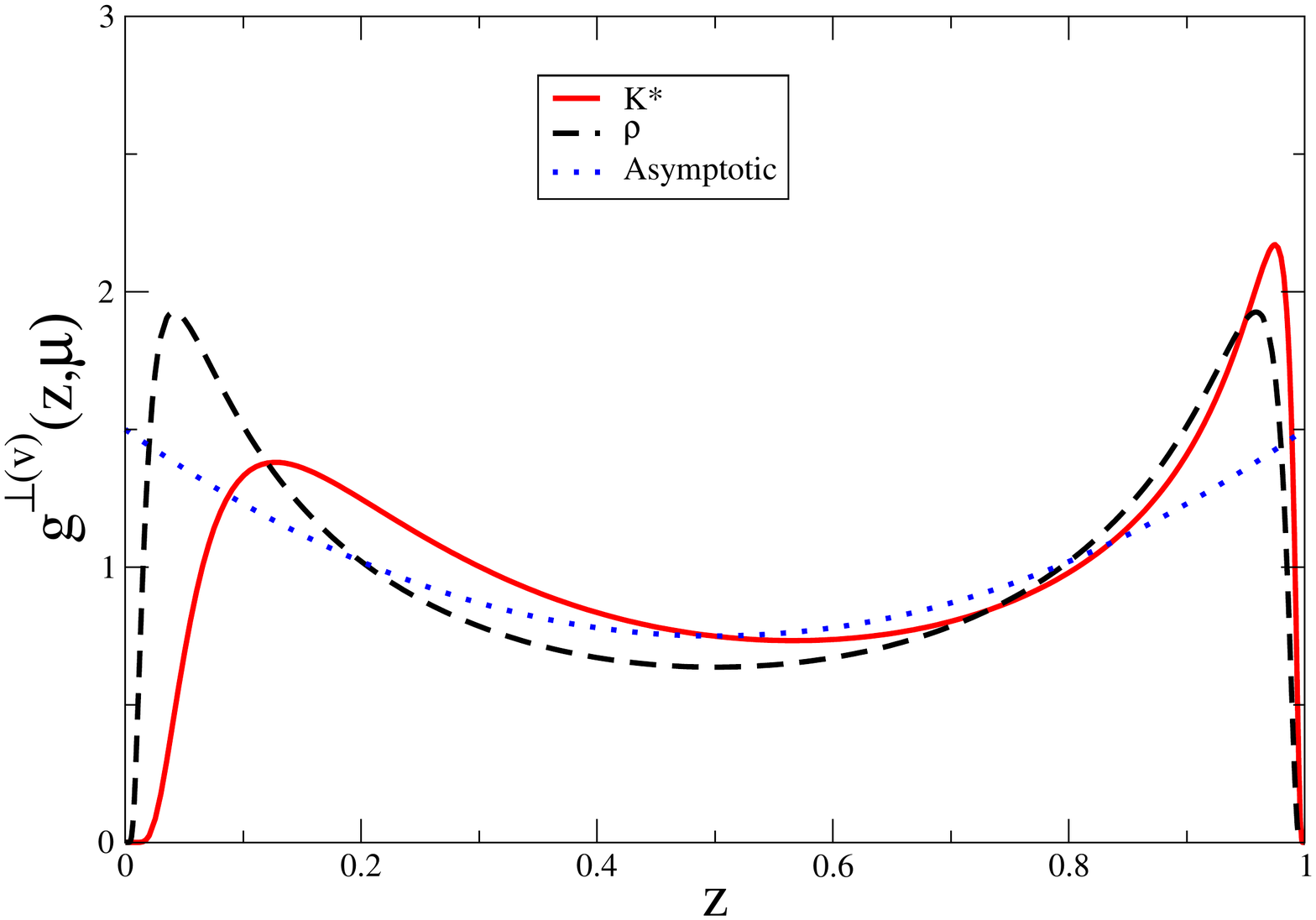}\includegraphics[height=.15\textheight]{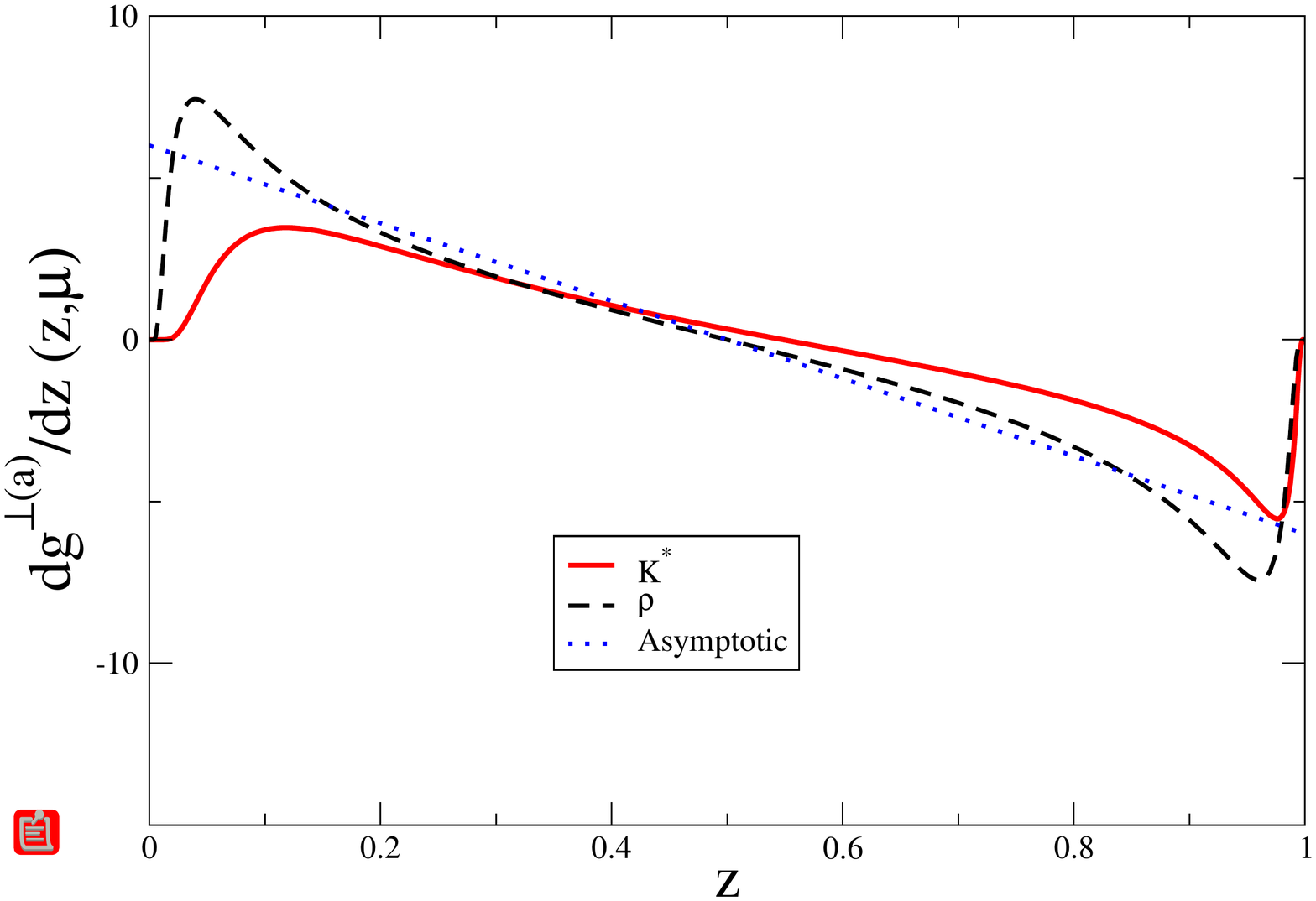}
\end{center}
  \caption{The twist-$2$ (left) DA, the vector twist-$3$ (middle) DA and the axial vector twist-$3$ (right) DA for for the transversely polarised vector meson. Solid Red: $K^*$, Dashed Black: $\rho$, Dotted Blue: Asymptotic.}
\label{fig:phiperp}
\end{figure}

The AdS/QCD DAs of a transversely polarised vector meson are related its AdS/QCD light-front wavefunction. In a standard notation, we have\cite{Ahmady:2012dy,Ahmady:2013cva}  
\begin{equation}
\phi_{V}^\perp(z,\mu) =\frac{N_c }{\pi f_{V}^{\perp}} \int \d
r \mu
J_1(\mu r) [m_q - z(m_q-m_{\bar{q}})] \frac{\phi_{V}^T(r,z)}{z(1-z)} \;,
\label{phiperp-phiT}
\end{equation}

\begin{equation}
g_{V}^{\perp(v)}(z,\mu)=\frac{N_c}{2 \pi f_{V} M_{V}} \int \d r \mu
J_1(\mu r)
\left[ (m_q - z(m_q-m_{\bar{q}}))^2 - (z^2+(1-z)^2) \nabla_r^2 \right] \frac{\phi_{V}^T(r,z)}{z^2 (1-z)^2
}
\label{gvperp-phiT}
\end{equation}
and
\begin{equation}
\frac{\d g_{V}^{\perp(a)}}{\d z}(z,\mu)=\frac{\sqrt{2} N_c}{\pi \tilde{f}_{V} M_{V}} \int \d r \mu
J_1(\mu r)
[(1-2z)(m_q^2- \nabla_r^2) + z^2(m_q+m_{\bar{q}})(m_q-m_{\bar{q}})]\frac{\phi_{V}^T(r,z)}{z^2 (1-z)^2} \;
\label{gaperp-phiT}
\end{equation}
where $\tilde{f}_V=f_V - f_V^{\perp} (m_q + m_{\bar{q}})/M_V$ with $f_V$ and $f_V^{\perp}$ being the decay constants of the meson. The wavefunction $\phi_{V}^{T}(r,z)$ is obtained by solving the holographic AdS/QCD Schroedinger equation of Brodsky and de Teramond \cite{deTeramond:2008ht}.\footnote{The resulting holographic wavefunction was recently used to generate successful predictions for the cross-sections of diffractive $\rho$ production \cite{Forshaw:2012im}.}  We find that these AdS/QCD DAs hardly depend on the renormalization scale $\mu$ for $\mu \ge 1$ GeV. They should therefore be viewed as parametrizations of the DAs at a low scale $\mu \sim 1$ GeV. In figure \ref{fig:phiperp}, we compare the twist-$2$ DA and twist-$3$ vector DA of the transversely polarised vector meson to their respective asymptotic forms ($\mu \to \infty$). 

\section{Branching ratio for $\bar{B}^{\circ} \to \rho^{\circ} \gamma$}

At leading power accuracy and next-to-leading accuracy in the strong coupling, the branching ratio for $\bar{B}^{\circ} \to \rho^{\circ} \gamma$ depends on two integrals involving the twist-$2$ DA namely
\begin{equation}
I_1^{\mbox{\tiny{tw2}}} (s_p,\mu_h) =\int_0^1 \d z \; h(s_p,\bar{z}) \phi_{\rho}^{\perp} (z,\mu_h) 
\label{I1}
\end{equation}
and
\begin{equation}
I_2^{\mbox{\tiny{tw2}}} (\mu_h) =  \int_0^1 \d z \; \frac{\phi_{\rho}^{\perp}(z,\mu_h)} {z} 
\label{I2}
\end{equation}
where $h(s_p,\bar{z})$\footnote{$s_p=(m_p/m_b)^2$ where $p=u,c$.}  is a hard scattering kernel and the hadronic scale $\mu_h=\sqrt{\Lambda_{\tiny{QCD}} m_b} \approx 2$  GeV. Beyond leading power accuracy, we compute the contributions from four annihilation diagrams\cite{Ahmady:2012dy}. The branching ratio then also depends on the twist-$3$ DA via two convolution integrals:
\begin{equation}
 I_{1}^{\mbox{\tiny{tw3}}} (\mu) = \int_0^1 \d z \; \frac{g_\rho^{\bot (v)}(z,\mu)}{zM_{B}^2+z\bar zM_\rho^2-m_q^2}
\label{I3}
\end{equation}
and
\begin{equation}
I_{2}^{\mbox{\tiny{tw3}}} (\mu) =\int_0^1 \d z \; \frac{zg_\rho^{\bot (v)}(z,\mu)}{zM_{B}^2+z\bar
  zM_\rho^2-m_q^2} \;.
\label{I4}
\end{equation}
In table \ref{tab:I1I2}, we show the values of the above integrals obtained when using either Sum Rules or AdS/QCD DAs. As can be seen, the AdS/QCD twist-$3$ avoids the divergence encountered with the corresponding Sum Rules DA. 
\begin{table}[h]
\begin{center}
\[
\begin{array}
[c]{|c|c|c|}
\hline
\mbox{Integral}&\mbox{SR} &\mbox{AdS/QCD} \\ \hline
I_1^{\mbox{\tiny{tw2}}}(s_c,\mu)&1.902+2.620 i  & 1.590+2.329 i \\ \hline
I_1^{\mbox{\tiny{tw2}}}(s_u,\mu)& -6.561+0.030 i&-8.866+0.027i  \\ \hline
I_1^{\mbox{\tiny{tw2}}}(0,\mu)& -6.660  & -8.989 \\ \hline
I_2^{\mbox{\tiny{tw2}}} & 3.330& 4.495 \\ \hline
I_{1}^{\mbox{\tiny{tw3}}}(\mu)& \infty  &   0.237 \\ \hline
I_{2}^{\mbox{\tiny{tw3}}}(\mu)& \infty & 0.036 \\ \hline
\end{array}
\]
\end{center}
\caption {AdS/QCD predictions at $\mu \sim 1$ GeV and SR predictions at a scale $\mu=2$ GeV for the two integrals contributing to the branching ratio of the decay $\bar{B}^{\circ} \to \rho^\circ \gamma$.}
\label{tab:I1I2}
\end{table}

The corresponding results for the branching ratio of the decay $\bar{B}^{\circ} \to \rho^{\circ} \gamma$ are shown in table \ref{tab:BRB0}. As can be seen, the power corrections to the branching ratio of the decay $\bar{B}^\circ \to \rho^\circ \gamma$ are numerically small so the end-point divergences arising from power corrections have no practical consequence when computing the branching ratio. The main source of uncertainty when predicting the branching ratio are therefore the hadronic uncertainties. 

\begin{table}[h]
\begin{center}
\textbf{Branching ratio ($\times 10^{-7}$) for $\bar{B}^\circ \to \rho^\circ \gamma$ }
\[
\begin{array}
[c]{|c|c|c|c|c|c|c|}
\hline
\mbox{DA}&\mbox{Accuracy}&\mbox{SR} &\mbox{AdS/QCD} & \mbox{PDG} & \mbox{Belle}  & \mbox{BaBar}\\ \hline
\mbox{tw}2 + \mbox{tw}3&\mbox{Lead.} (\alpha_s^1)+ \mbox{Anni.}[\alpha_s^0, (1/m_b)^2] &  &7.67&8.6 \pm 1.5  &7.8 \pm^{1.7}_{1.6} \pm^{0.9}_{1.0}& 9.7\pm^{2.4}_{2.2} \pm^{0.6}_{0.6}\\ \hline
\mbox{tw}2 &\mbox{Lead.}(\alpha_s^1) + \mbox{Anni.}[\alpha_s^0,(1/m_b)] & 7.86&7.65 & & & \\ \hline
\end{array}
\]
\end{center}
\caption {Sum Rules and AdS/QCD predictions for the branching ratio ($\times 10^{-7}$) of $\bar{B}^\circ \to \rho^\circ \gamma$ using AdS/QCD or Sum Rules compared to the measurements from Belle \cite{Taniguchi:2008ty}, BaBar \cite{Aubert:2008al} and the average value from PDG \cite{Beringer:1900zz}.} 
\label{tab:BRB0}
\end{table}

\section{The isospin asymmetry in $B \to K^* \gamma$}
Being given in terms of ratios of branching ratios, the isospin asymmetry is less sensitive to the hadronic uncertainties. However,  in the decay $B \to K^* \gamma$, the isospin asymmetry is a power correction, i.e. it vanishes to leading power accuracy.  Therefore, it is important to address the end-point divergence problem when computing this asymmetry. The isospin asymmetry depends on four convolution integrals namely\cite{Kagan:2001zk}
 \begin{equation}
F_{\perp} (\mu_h)=\int_0^1 \d z \;  \frac{\phi^{\perp}_{K^*}(z,\mu_h)}{3(1-z)}
\label{Fperp}
\end{equation}
\begin{equation}
G_{\perp}(s_c,\mu_h)=\int_0^1 \d z \; \frac{\phi^{\perp}_{K^*}(z, \mu_h)}{3(1-z)} G(s_c,\bar{z})
\label{Gperp}
\end{equation}
\begin{equation}
X_{\perp}( \mu_h) = \int_0^1 \d z \; \phi_{K^*}^{\perp}(z, \mu_h) \left( \frac{1 + \bar{z}}{3 \bar{z}^2} \right)
\label{Xperp}
\end{equation}
and
\begin{equation}
H_{\perp}(s_c, \mu_h) = \int_0^1 \d z \; \left(g^{\perp(v)}_{K^*} (z,\mu_h) - \frac{1}{4}\frac{\d g_{K^*}^{\perp (a)}}{\d z} (z,\mu_h)\right)G(s_c,\bar{z})
\label{Hperp}
\end{equation}
where $G(s_c,\bar{z})$ is the penguin function \cite{Kagan:2001zk} with $\bar{z}=1-z$.  The first three integrals $F_{\perp}$, $G_{\perp}$ and $X_{\perp}$ depend on  the twist-$2$ DA  while $H_{\perp}$ depends  on the twist-$3$ DAs. 
\begin{table}[h]
\begin{center}
\[
\begin{array}
[c]{|c|c|c|}\hline \mbox{Integral} & \mbox{SR} &\mbox{AdS/QCD} \\
\hline X_{\perp} & \infty &  26.9\\ 
\hline F_{\perp} & 1.14& 1.38\\
\hline G_{\perp} &2.55 + 0.43 i & 2.89 + 0.30 i\\ 
\hline H_{\perp} &2.48 + 0.50 i & 2.12 + 0.21 i\\ \hline
\end{array}
\]
\end{center}
\caption {Predictions for the four convolution integrals contributing to the isospin asymmetry in the decay $B \to K^* \gamma$ using Sum Rules  DAs at a scale $\mu=2$ GeV and the AdS/QCD DAs at a scale $\mu \sim 1$ GeV.}
\label{tab:integrals}
\end{table}
As can be seen in table \ref{tab:integrals}, the twist-$2$ AdS/QCD DA avoids the divergence encountered with the corresponding Sum Rules DA. With the AdS/QCD DAs, we predict  an isospin asymmetry of $3.3\%$ in agreement with the most recent average measured value of $(5.2 \pm 2.6)\%$ quoted by the Particle Data Group. 
\bibliographystyle{JHEP}
\bibliography{RSandapenMA}
\end{document}